\documentclass[aps,pra,twocolumn,showpacs,floatfix]{revtex4}

\usepackage{hyperref}

\usepackage{amstext}
\usepackage{amsmath}

\usepackage{theorem}
\theoremheaderfont{\scshape}
\newtheorem{theorem}{Proposition}

\usepackage{graphicx}

\newcommand{\sx}{\hat \sigma^{(x)}}
\newcommand{\sy}{\hat \sigma^{(y)}}
\newcommand{\sz}{\hat \sigma^{(z)}}
\newcommand{\Sp}{\hat \sigma^{(+)}}
\newcommand{\Sm}{\hat \sigma^{(-)}}

\newcommand{\tr}{\mathop{\mbox{tr}}\nolimits}
\newcommand{\tensorprod}{\mathop{\otimes}\limits}
\def\bra#1{\mathinner{\langle{#1}|}}
\def\ket#1{\mathinner{|{#1}\rangle}}

\def\Bra#1{\left<#1\right|}
\def\Ket#1{\left|#1\right>}

\begin{document}

\title{Direct versus measurement assisted bipartite entanglement in multi-qubit systems and their dynamical generation in spin systems}

\author{M\'aty\'as Koniorczyk}
\altaffiliation{On leave from: Research Institute for Solid State Physics and Optics of the Hungarian Academy of Sciences, Budapest}
\affiliation{Research Centre for Quantum Information,
Institute of Physics, Slovak Academy of Sciences
D\'{u}bravsk\'{a} cesta 9,
845 11 Bratislava, Slovakia}

\author{Peter Rap\v can}
\affiliation{Research Centre for Quantum Information,
Institute of Physics, Slovak Academy of Sciences
D\'{u}bravsk\'{a} cesta 9,
845 11 Bratislava, Slovakia}

\author{Vladim\' \i r Bu\v zek}
\affiliation{Research Centre for Quantum Information,
Institute of Physics, Slovak Academy of Sciences
D\'{u}bravsk\'{a} cesta 9,
845 11 Bratislava, Slovakia}
\affiliation{QUNIVERSE, L\'\i \v s\v cie \'udolie 116, 841 04
  Bratislava, Slovakia}

\date{March 14, 2005.}

\begin{abstract}
  We consider multi-qubit systems and relate quantitatively the
  problems of generating cluster states with high value of concurrence
  of assistance, and that of generating states with maximal bipartite
  entanglement. We prove an upper bound for the concurrence of
  assistance.  We consider dynamics of spin-1/2 systems that model
  qubits, with different couplings and possible presence of magnetic
  field to investigate the appearance of the discussed entanglement
  properties.  We find that states with maximal bipartite entanglement
  can be generated by an XY Hamiltonian, and their generation can be
  controlled by the initial state of one of the spins. The same
  Hamiltonian is capable of creating states with high concurrence of
  assistance with suitably chosen initial state.  We show that the
  production of graph states using the Ising Hamiltonian is
  controllable via a single-qubit rotation of one spin-1/2 subsystem
  in the initial multi-qubit state.  We shown that the property of
  Ising dynamics to convert a product state basis into a special
  maximally entangled basis is temporally enhanced by the application
  of a suitable magnetic field. Similar basis transformations are
  found to be feasible in the case of isotropic XY couplings with
  magnetic field.
\end{abstract}

\maketitle

\section{Introduction}

One of the key features of a physical system for quantum information
processing (QIP) is quantum entanglement. The problem of entanglement
of multipartite systems is far from being completely understood, and
it has numerous interesting aspects.

One of the possible approaches to multipartite entanglement is to
search for quantum states with prescribed bipartite entanglement
properties~\cite{KoashiBI00,PleschB03,PleschB02}. This is a nontrivial
task as there exist limitations on the bipartite entanglement in
multipartite systems, which were quantified by Coffmann, Kundu and
Wootters~\cite{CoffmanKW00}. In a pioneering work, O'Connor and
Wootters~\cite{OConnorW01} have considered a system of quantum bits,
and have searched for an entangled state of these with maximal
bipartite entanglement. This state appears to be the ground state of
the antiferromagnetic Ising model, the spins representing the qubits.
This illustrates the relation between states of maximal bipartite
entanglement and the spin couplings known from statistical physics. We
will refer to this approach as the question of \emph{direct bipartite
  entanglement}, as the relevant quantity is the bipartite
entanglement present in the system as it is.

Another approach to the problem of multipartite entanglement is
related to cluster~\cite{BriegelR01} and graph~\cite{HeinEB04} states.
These are genuine multipartite entangled states, which can be
projected onto a maximally entangled state of any chosen two spins by
a von Neumann measurement on the others. Such states arise dynamically
in a system of spins with pairwise Ising couplings. They constitute
the fundamental entangled resource for one-way quantum
computers~\cite{RaussendorfB01,RaussendorfBB03}. It is an interesting
property of the Ising dynamics in this case, that it transforms a
whole basis of product states into a basis which consists of cluster
or graph states.  In this way a basis transformation from a product
state basis to a special -- in a sense maximally -- entangled basis is
realized.

These states are the starting points for the second approach, the
bipartite entanglement in multipartite systems available via assistive
measurements on all but two subsystems. The two key concepts in its
quantitative description are entanglement of
assistance~\cite{DiVincensoFMSTU} (or concurrence of
assistance~\cite{LaustsenVV03}, quantifying the entanglement available
via assistive measurements, and localizable
entanglement~\cite{VerstraetePC04b,quantph0411123}. The computational
feasibility of concurrence of assistance for a pair of qubits makes
the quantitative study of a part of this question feasible.

One of our aims is to relate the above two approaches. We will show
that the optimizations of direct and measurement assisted bipartite
entanglement are indeed related. Our other task is to study these
generic features in actual spin systems, as such systems do appear
quite naturally in this context.

Coupled spin systems have attracted a vast amount of research interest
in the quantum information community recently. The couplings studied
in statistical physics allow for performing certain tasks in QIP such
as e.g. quantum state
transfer~\cite{Bose03,ChristandlDEL04,OsborneN04}, realization of
quantum gates~\cite{SchuchS03,YungLB04}, and quantum
cloning~\cite{ChiaraFMMM04}. As the systems of coupled spins are
appropriate models for solid state systems, and also for quantum
states in optical lattices in certain cases~\cite{Garcia-RipollC03},
they bear actual practical relevance.

In the second part of this paper we focus on dynamical generation of
entanglement. We consider a system initially in a pure product state,
and investigate the entanglement of the states of the system
throughout the evolution.  The ``prototype'' of such entanglement
generation is that of cluster and graph states. The various aspects of
the dynamical behavior of entanglement in spin systems has been
considered by several authors
recently~\cite{AmicoOPRP04,Subrahmanyam04a,PlastinaAOF04,quantph0409039,quantph0409048,VidalPA04}.

In addition to interpolating between the two approaches to bipartite
entanglement in multipartite systems, we consider the possibility of
controlling the process through the initial state of the system.  We
address the following question. Is it possible to dynamically generate
states with optimal direct bipartite entanglement? We find a positive
answer, and also that the same couplings are capable of producing
states with high bipartite entanglement available via measurements, if
a different initial state is chosen. Our main tool of describing
measurement assisted bipartite entanglement will be concurrence of
assistance. We will examine the possibility of controlling the
behavior of this entanglement generation by the initial state of the
system.  This is analogous to the control of quantum operations in
programmable quantum
circuits~\cite{quantph0102037,prl79_321,pra65_022301,pra66_042302}.
Finally we show that a suitably chosen magnetic field can enable
couplings different from Ising to create whole entangled bases
resembling those of cluster states regarding concurrence of
assistance. (Note that the generation of cluster states with non-Ising
couplings was considered very recently in Ref.~\cite{quantph0410145})
In addition, the application of magnetic field in the case of Ising
couplings can temporally enhance the presence of high pairwise
concurrence of assistance.

As we are mainly interested in illustrating generic features and
certain examples of entanglement behavior, a part of our results
concerning actual spin systems is simply computed by numerical
diagonalization of the appropriate Hamiltonians, even though we
present some analytical considerations where we find them appropriate.
Thus some of our considerations are limited to an order of 10 spins,
even though according to the numerical experience, they seem to be
scalable. This number coincides with that of the quantum bits expected
to be available in quantum computers in the near future. As the
realization of the discussed couplings is not necessarily restricted
to spins, our results may become directly applicable in such systems.
We consider two topologies of the pairwise interactions: a \emph{ring}
where each spin interacts with its two neighbors, and also the
\emph{star} topology where the interaction is mediated by a central
spin interacting with all the others. This was found interesting from
the point of view of entanglement distribution~\cite{HuttonB04} and
also from other aspects of its dynamics~\cite{BreuerBP04} recently.

The paper is organized as follows: in the introductory
Section~\ref{sect:entangmeas} we briefly review the entanglement
measures we use in the following. Section~\ref{sect:graphstates} is
devoted to the review of the dynamical generation of cluster and graph
states in spin systems, which is the background of the second part of
the paper. In Section~\ref{sect:upb} we present two interesting
properties of concurrence of assistance, which relates the two above
mentioned approaches to bipartite entanglement in multipartite systems,
and will be useful in the following. In Section~\ref{sec:control}, the
controlled generation of specific entangled states is addressed.
Section~\ref{sect:bases} is devoted to the enhanced generation of
certain entangled bases with the help of magnetic field.
Section~\ref{sect:concl} summarizes our results.

\section{Entanglement measures}
\label{sect:entangmeas}

In this Section we give an overview in a nutshell of the entanglement
measures and related quantities that will be used throughout this
paper.

\paragraph{One-tangle.}

For a bipartite system $A\bar{A}$ (A being a qubit, $\bar{A}$ being
the rest of the system) in the pure state $\Ket{\Psi}_{A\bar{A}}$, the
one-tangle~\cite{HillW97} of either of the subsystems
\begin{equation}
T\left(\Ket{\Psi}_{A{\bar{A}}}\right)=
4\det(\varrho_{A})
\label{eq:entanglement}
\end{equation}
(where $\varrho_{A}=\tr_{\bar{A}}\Ket{\Psi}_{A\bar{A}}\Bra{\Psi}$), is a
measure of entanglement. It quantifies the entanglement between the
qubit $A$ and the rest of the system, including all multipartite
entanglement between qubit A and the sets all the subsystems in
$\bar{A}$.

Although there is an extension of one-tangle to mixed states, it is
not computationally feasible except for the case of 2 qubits, in which
case one-tangle is equal to the square of concurrence. This justifies
the following interpretation: the square root of one-tangle is the
concurrence of such a two-qubit system in a pure state, for which the
density matrix of one of the qubits is equal to that of qubit A. This
means, it would be the concurrence itself if the subsystem $\bar{A}$
were also a qubit.

\paragraph{Concurrence.}

Having a bipartite system in a mixed state, a way of defining their
entanglement is to consider the average entanglement of all the pure
state decompositions of the state. This quantity is termed as the
\emph{entanglement of formation}:
\begin{equation}
  E(\varrho)=\min\sum_{i}p_{i}E(\Ket{\Psi_{i}}),\quad\text{so
    that}\,\,\sum_{i}p_{i}\Ket{\Psi_{i}}\Bra{\Psi_{i}}=\varrho.
\label{eq:entform}
\end{equation}
This is a kind of generalization of the entanglement defined in
Eq.~\eqref{eq:entanglement}.  Its additivity is one of the most
interesting open questions of QIT.

The definition of entanglement of formation supports the following
interpretation: imagine that the bipartite system as a whole is a
subsystem of a large system. Entanglement of formation measures the
bipartite entanglement available on average if everything but the
bipartite subsystem is simply dropped.

If the system in argument consists of two qubits, there is a closed
form for entanglement of formation found by
Wootters~\cite{Wootters98}.  This consideration includes another
entanglement measure.

Given the two-qubit density matrix $\varrho$, one calculates the
matrix \begin{equation}
  \tilde{\varrho}=(\sigma^{(y)}\otimes\sigma^{(y)})\varrho^{*}(\sigma^{(y)}\otimes\sigma^{(y)}),
\label{eq:wootterstilde}
\end{equation}
where $*$ stands for complex conjugation in the product-state basis.
$\tilde{\rho}$ describes a very unphysical state for an entangled
state, while it is a density matrix for product states.

In the next step one calculates the eigenvalues $\lambda_{i}$
($i=1\ldots4$) of the Hermitian matrix
\begin{equation}
\label{eq:rhomatrix}
  \hat{R}=\sqrt{\sqrt{\varrho}\tilde{\varrho}\sqrt{\varrho}},
\end{equation}
which are in fact square roots of the eigenvalues of the non-Hermitian
matrix
\begin{equation}
  \hat{R}_{2}=\varrho\tilde{\varrho}.
\label{eq:R2}
\end{equation} 
Concurrence is then defined as
\begin{equation}
  C(\varrho)=\max(0,\lambda_{1}-\lambda_{2}-\lambda_{3}-\lambda_{4}),
\label{eq:concurrence}
\end{equation}
where the eigenvalues are put into a decreasing order.
Entanglement of formation is a monotonously increasing function of
concurrence:
\begin{eqnarray}
   E(\varrho)=h\left(\frac{1+\sqrt{1-C(\varrho)^{2}}}{2}\right),\nonumber \\
 h(x):=-x\log_{2}(x)-(1-x)\log_{2}(1-x).
\end{eqnarray}
Thus concurrence can be used as an entanglement measure on its own
right.

In multipartite systems the one-tangle and concurrence are linked by
the Coffmann-Kundu-Wootters inequalities
\begin{equation}
  \label{eq:CKW}
  T_k \geq \sum\limits_{l\neq k} C_{kl}^2
\end{equation}
which have be proven initially for three qubits in a pure state and
certain classes of multi-qubit states. For a long time they were
conjectured to be true in general. This conjecture was very recently
proven~\cite{quantph0502176}. These inequalities set limitations to
the bipartite entanglement that can be present in a multipartite
system.

\paragraph{Concurrence of assistance.}

Consider again a bipartite system described by the density operator
$\varrho$. One can follow a route complementary to that in case of
entanglement of formation and ask what is the \emph{maximum} average
entanglement available amongst the pure state realizations, termed as
the \emph{entanglement of assistance}~\cite{Wootters98}:
\begin{eqnarray}
E_{\text{assist}}(\varrho)=\max\sum_{i}p_{i}E(\Ket{\Psi_{i}}),
\nonumber \\ 
\text{so that}\,\,\sum_{i}p_{i}\Ket{\Psi_{i}}\Bra{\Psi_{i}}=\varrho,&
\label{eq:entass}
\end{eqnarray}
c.f. Eq.~\eqref{eq:entform}.

Interpreting again the bipartite system as a subsystem of a larger
system, one can consider that the whole system is in a pure state,
that is, we have a purification of $\varrho$ at hand. In this case
entanglement of assistance describes the maximum entanglement
available on average in the bipartite system, when a collaborating
third party, instead of omitting the rest of the system as in the case
of entanglement of formation, makes optimal von Neumann measurements
on it. Although entanglement of assistance is not an entanglement
measure according to some definitions, it is a very informative
quantity regarding entanglement.

Having a system of two qubits, one can also use concurrence instead of
entanglement in Eq.~\eqref{eq:entass}, yielding the definition of
\emph{concurrence of assistance}:
\begin{eqnarray}
C_{\text{assist}}(\varrho)=\max\sum_{i}p_{i}C(\Ket{\Psi_{i}}\Bra{\Psi_{i}}),
\nonumber \\ 
\text{so that}\,\,\sum_{i}p_{i}\Ket{\Psi_{i}}\Bra{\Psi_{i}}=\varrho.&
\label{eq:concass}
\end{eqnarray}
 The advantage of this quantity is, that it can be easily calculated
for two qubits. As it is shown in~\cite{LaustsenVV03}, it is simply
\begin{equation}
C_{\text{assist}}(\varrho)=
\tr\sqrt{\sqrt{\varrho}\tilde{\varrho}\sqrt{\varrho}}=
\sum_{i=1}^{4}\lambda_{i},
\label{eq:cassist}
\end{equation}
c.f. Eq.~\eqref{eq:concurrence}. Note that this quantity is
essentially a fidelity between the physical density matrix $\varrho$
and the matrix $\tilde{\varrho}$, which is physical for separable
states only.

Thanks to the formula in Eq.~\eqref{eq:cassist}, concurrence of
assistance is not only an informative quantity, but it is as feasible
as concurrence itself in the case of qubit pairs.

\section{Graph states revisited}
\label{sect:graphstates}

In this Section we briefly review the properties of the Ising dynamics
for spin-1/2 particles without magnetic field, which are known from
Refs.~\cite{BriegelR01,HeinEB04}.  We will talk about spins in this
context, and the $\sz$ eigenstates will represent the computational
basis: $\ket{0}=\ket{\uparrow}$, $\ket{1}=\ket{\downarrow}$. Consider a
set of spins, with pairwise interactions between them:
\begin{equation}
  \label{eq:IsingnoB}
  \hat H = -\sum\limits_{\langle k,l \rangle}
  \sx_k \otimes \sx_l
\end{equation}
where the summation ${\langle k,l \rangle}$ goes over those spins
which interact with each other. (Hence the name graph states for the
states to be considered here: the geometry can be envisaged as a
graph, where the vertices are the spins, and the edges represent
pairwise Ising interactions.) As the summands in
Eq.~\eqref{eq:IsingnoB} commute, the time evolution can be written as
a product of two-spin unitaries
\begin{equation}
  \label{eq:U}
  \hat U(\tau) =e^{-i\hat H \tau}=
  \prod\limits_{\langle k,l \rangle}
  \hat U_{k,l}(\tau),
\end{equation}
where
\begin{equation}
  \label{eq:Ukltau}
  \hat U_{k,l}(\tau)=e^{i \sx_k \otimes \sx_l  \tau}.
\end{equation}
Here $\tau$ stands for the scaled time measured in arbitrary units.

First we study the time instant $\tau=\frac{\pi}{4}$: one may directly
verify that
\begin{equation}
  \label{eq:Ukl}
    \hat U_{k,l}=
    \frac{1}{\sqrt{2}}
    \left(
      \hat 1 +i \sx _k \otimes \sx _l
    \right).
\end{equation}
The evolution operators without a time argument will denote those for
$\tau=\frac{\pi}{4}$ in what follows.  These describe conditional
phase gates in a suitably chosen basis. Let us assume that the system
is initially in a state $\ket{e_m}$ of the computational basis, a
common eigenvector of all the $\sz$-s:
\begin{equation}
  \label{eq:szeig}
  \sz_n \ket{e_m} = e_{n,m} \ket{e_m}, \qquad e_{n,m}=\pm 1.
\end{equation}
The state $\hat U \ket{e_m}$ will be an eigenvector of the following
complete set of commuting observables:
\begin{equation}
  \label{eq:K}
  \hat K_n=\hat U \sz _n \hat U^\dag,
\end{equation}
with the same eigenvalues as the $e_{n,m}$-s in Eq.~\eqref{eq:szeig}.
The operators $\hat K_n$ in Eqs.~\eqref{eq:K} depend on the geometry of
the graph.  They can be evaluated simply by utilizing the following
relations:
\begin{eqnarray}
  \label{eq:Ucomm}
  \hat U_{k,l} \sx _k \hat U_{k,l}^\dag &=&  \sx _k \nonumber \\
  \hat U_{k,l} \sy _k \hat U_{k,l}^\dag &=& -\sz _k \otimes \sx _l
\nonumber \\
  \hat U_{k,l} \sz _k \hat U_{k,l}^\dag &=& \sy _k \otimes \sx _l
\nonumber \\
  \hat U_{k,l} \hat \sigma^{(x,y,z)} _m \hat U_{k,l}^\dag &=&
\sigma_m^{(x,y,z)}
  \quad (m\neq k,l).
\end{eqnarray}
which can be verified directly by substituting Eq.~\eqref{eq:Ukl} into
Eq.~\eqref{eq:K}. The joint eigenstates of these operators are termed
as \emph{graph states}~\cite{HeinEB04}. It can be shown that many of
the so arising states corresponding to different graphs are local
unitary equivalent.

As an example, consider a ring of $N$ spins with pairwise Ising
interaction. In this case
\begin{equation}
  \label{eq:Kring}
\hat  K^{\text{(ring)}}_l=
-\sx_{l-1}\otimes \sz_{l} \otimes \sx_{l+1},
\end{equation}
where the arithmetics in the indices is understood in the modulo N
sense. The common eigenstates of these commuting variables are termed
as \emph{cluster states}, and they were introduced in
Ref.~\cite{BriegelR01}, although in a different basis. They are
suitable as an entangled resource for one-way quantum
computers~\cite{RaussendorfB01}.

Note that $\hat U(\pi)=-\hat 1$ in general. Specially for a ring
topology, $\hat U(\pi/2)=-\hat 1$ holds too. This means that the
evolution is periodic: at such time instants the initial state appears
again, which is a computational basis state. Thus the Ising dynamics
without magnetic field produces oscillations between the computational
basis state and a graph (or in some of the cases, cluster) state. The
achieved graph state is selected by the initial basis state.

\begin{figure}[htbp]
  \centering
  \includegraphics{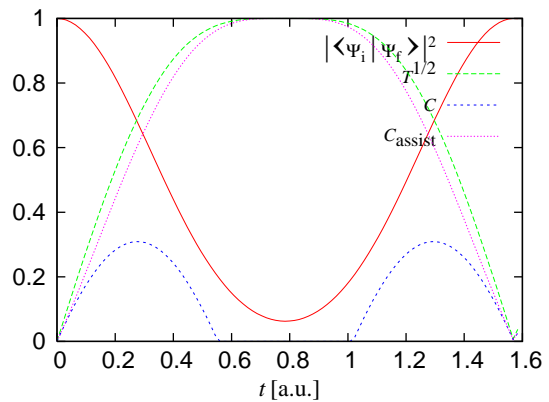}
  \caption{(Color online.) 
    Overlap with the initial state and entanglement measures 
    for the first two qubits, during the entanglement oscillations for
    five spins in a ring, generated by the Ising Hamiltonian without
    magnetic field in~\eqref{eq:IsingnoB}.  In the initial state all
    spins are up, thus in state $\ket{0}$ if we consider qubits. The
    plotted quantities are dimensionless.}
  \label{fig:entangosc}
\end{figure}

To obtain a more complete picture on the whole process of the
entanglement oscillations, we plot the temporal behavior of the
entanglement quantities in Fig.~\ref{fig:entangosc} for the ring
topology.
In the figure we observe that the concurrence of assistance of the
qubit pair is almost equal to the square root of one-tangle of one of
the constituent spins. We will show later in this paper that the
square root of one tangle is an upper bound for concurrence of
assistance. Thus for the states in argument, the entanglement of a
subsystem with the rest of the system can be indeed ``focused'' to a
pair of qubits via suitably chosen measurement on the rest of the
system. This is obvious for the cluster states, but it appears to hold
for the most of the time evolution.

The dynamical entanglement behavior of the systems in argument can be
controlled by the appropriate choice of the initial state. Consider
for instance the following polarized initial state:
\begin{equation}
  \label{eq:instate_Ising}
  \Ket{\Psi_{\text{A}}(t=0)} = 
    \tensorprod_{k=1}^N
          \left(   
             \cos \left(\frac{\theta}{2}\right) \ket{0}_k + 
             \sin \left(\frac{\theta}{2}\right) \ket{1}_k
          \right).
\end{equation}
The ``A''index reflects that \emph{all} the spins are rotated from the
$z$ direction in the same way. This state can be prepared by a
simultaneous one-qubit rotation, which is available even in optical
lattice systems. If $\theta=l\pi$ ($l$ being integer), we obtain the
graph state periodically, while for $\theta=l\pi/2$ the state is
stationary, thus no entanglement will be generated. Between these
values, the entanglement measured by one-tangle or concurrence of
assistance is a monotonous and continuous function of $\theta$ for all
values of time.  Thus by varying this parameter of the initial state,
one can control the amount of the generated entanglement.

From the above discussion we find that Ising dynamics without magnetic
field has the following properties from the point of view of
entanglement generation:

\begin{enumerate}
\item The generated bipartite entanglement is always small.
\item In the case of the cluster states one can project the state with
  certainty to a maximally entangled pair of two spins by a
  measurement on the others. Moreover, required measurement is a local
  one.
\item \emph{All} the states of the computational basis are
  periodically transferred into states which have properties 1-2.
\item One can control the amount of the dynamically generated
  entanglement by a parameter of the initial state, which can be
  altered by the same local rotation applied on all the spins.
\end{enumerate}

During our investigations we will check which of these properties may
arise under different couplings, initial states and topologies.

\section{Two properties of concurrence of assistance}
\label{sect:upb}

In this Section we present two properties of concurrence of
assistance for multi-qubit systems.

Our first proposition formulates an upper bound of concurrence of
assistance.
\begin{theorem}
\label{thm:upb}
For an arbitrary state of two qubits $A$ and $B$, square root of the
one-tangle of either qubits serves as an upper bound for concurrence
of assistance, i.e.:
  \begin{equation}
    \label{eq:lemmst}
  \sqrt{T_A}\geq  C^{\text{assist}}_{AB}.
\end{equation}
\end{theorem}
Proof: Consider the ensemble realization of the state $\varrho_{AB}$
of the qubits A,B
\begin{equation}
\label{eq:bndpr1}
  \varrho_{AB}=\sum_k p_k \ket{\xi_k} \bra{\xi_k}
\end{equation}
which provides the maximum in
Eq.~\eqref{eq:concass}, and use the notation
\begin{equation}
  \label{eq:rhok}
  \varrho_k=\tr_B \ket{\xi_k} \bra{\xi_k},
\end{equation}
thus
\begin{equation}
  \label{eq:rhoa}
   \varrho_{A}=\tr_B \varrho_{AB}=\sum_k p_k\varrho_k,
\end{equation}
due to the linearity of the partial trace.  Substituting
Eq.~\eqref{eq:rhoa} into the definition in Eq.~\eqref{eq:entanglement}
we obtain
\begin{equation}
  \label{eq:sqt}
  \sqrt{T_A}=2\sqrt{\det\left( \sum_k p_k \varrho_k\right)},
\end{equation}
while according to the definition in Eq.~\eqref{eq:concass},
\begin{equation}
  \label{eq:cassp}
  C^{\text{assist}}_{AB}=2\sum_k \sqrt{\det(p_k\rho_k)},
\end{equation}
where we have exploited the fact that for pure states
\begin{equation}
  C( \ket{\xi_k})=2\sqrt{\det \varrho_k}.
\end{equation}
Substituting Eqs.~\eqref{eq:sqt} and ~\eqref{eq:cassp} into the
statement of the Proposition in inequality~\eqref{eq:lemmst}, what we
have to show is that
\begin{equation}
  \sum_k \sqrt{\det(p_k\varrho_k)} \leq 
  \sqrt{\det\left( \sum_k p_k \varrho_k\right)}.
\end{equation}
This is a consequence of the recursive application of the
inequality~\eqref{eq:mainineq}, which is proven in
Appendix~\ref{app:ineqproof}.  \hfill QED.

Intuitively, in the spirit of the considerations concerning lower
bound of localizable entanglement in Ref.~\cite{quantph0411123}, we can
claim that a local measurement on the ancillary systems of a
purification of $\varrho_{AB}$ cannot create additional entanglement
between the spin $A$ and the rest of the system $\bar{A}$, as such a
measurement is an operation on the complementary system. Thus, by
choosing the optimal measurement we can, at best, concentrate all of
the originally available entanglement ($\sqrt{T_{A}}$) into the
entanglement between the qubits $A$ and $B$.

The appearance of the one-tangle in the context of concurrence of
assistance suggests that there might be some relation with the CKW
inequalities, and this is the case indeed. Nevertheless, it is simple
to prove the following:
\begin{theorem}
  \label{thm:ckw}
  For a system of three qubits $A$,$B$,$C$ in a pure state, 
  \begin{equation}
    C_{AB}=C^{\text{assist}}_{AB}\ {\mathrm{and}}\  
    C_{AC}=C^{\text{assist}}_{AC}
  \end{equation}
  implies that the Coffmann-Kundu-Wootters inequalities in
  Eq.~\eqref{eq:CKW} are saturated, thus
  \begin{equation}
    C_{AB}^2+C_{AC}^2=T_A
  \end{equation}
  holds
\end{theorem}
This immediately follows from the same derivation as in
Ref.~\cite{CoffmanKW00} by exploiting the fact that the matrices $R_2$
of Eq.~\eqref{eq:R2} for subsystems $AB$ and $AC$ have rank one due to
the conditions of the proposition. (C.f. Eqs.~\eqref{eq:concurrence}
and~\eqref{eq:cassist}).

Proposition~\ref{thm:ckw} relates the direct and measurement assisted
approach to bipartite entanglement in multipartite systems.  The
question remains open, of course, whether it is true for more
parties, too.

As already pointed out in Section~\ref{sect:graphstates}, for the
graph states themselves $\sqrt{T_A} = C^{\text{assist}}_{AB}=1$, and
besides $\sqrt{T_A} \approx C^{\text{assist}}_{AB}$ holds throughout
the whole time evolution generated by Ising couplings. According to
Proposition~\ref{thm:upb} it is correct to call such states as those with
maximal concurrence of assistance.  Meanwhile $C_{AB}\ll
C^{\text{assist}}_{AB}$, which suggests that CKW inequalities are far
from being saturated, which is indeed the case. The generated
entanglement is essentially multipartite, but it can be converted to
bipartite via a measurement. On the other hand, if CKW inequalities
are saturated, then we can expect concurrence of assistance being
below the square-root of one-tangle. Besides, the question naturally
arises, whether it is possible to dynamically create entanglement
oscillations in spin systems which saturate CKW inequalities instead.

\section{Controlled generation of concurrence and concurrence of assistance}
\label{sec:control}

Now we turn our attention to spin-1/2 systems as those naturally
realize multi-qubit systems. We assign the $\sz$ eigenstates as the
computational basis states as $\ket{0}=\ket{\uparrow}$,
$\ket{1}=\ket{\downarrow}$. We will use the qubit notation for
simplicity.

We have seen in Section~\ref{sect:graphstates} that certain states
with maximal concurrence of assistance can be generated in dynamical
oscillations, and the control over the available entanglement is
realized by the altering of the initial state.  This control requires
a simultaneous operation on all the spins, and as for bipartite
entanglement, it effects the entanglement available via assistive
measurements only, as concurrence itself takes low values throughout
the evolution. First we consider whether it is possible to control the
concurrence itself too, and if it is possible to control the evolution
by varying a single spin only.

Consider first a system of $N+1$ spins with XY couplings:
\begin{equation}
\hat H_{XY}=-\sum\limits_{<i,j>} \sx_i \sx_j + \sy_i \sy_j,
\label{eq:XYnoB}
\end{equation}
in a star topology: spin $0$ is the middle one, while spins $1$ to $N$
are the outer ones, each coupled to the central one. Even though the
summands of the Hamiltonian do not commute, the eigenvalues and
eigenvectors can be calculated. One would expect that the state of the
middle spin can control the entanglement behavior, as the interaction
of the outer spins is mediated by this one. Indeed, if one considers
the initial state where only the middle spin is rotated, the others
point upwards, i.e. they are in the state $\ket{0}$:
\begin{widetext}
\begin{equation}
  \label{eq:inXY}
    \Ket{\Psi_{\text{M}}(t=0)}=
   \left(\cos \left(\frac{\theta}{2}\right) \ket{0}_0 + 
   \sin \left(\frac{\theta}{2}\right) \ket{1}_0 \right)
   \otimes 
   \tensorprod_{k=1}^N \ket{0}_k,
\end{equation}
the time evolution, as shown in
Appendix~\ref{app:andyn}, reads 
\begin{eqnarray}
  \label{eq:XYtime}
    \Ket{\Psi_{\text{M}}(t)}=
   &\cos \left(\frac{\theta}{2}\right)& 
   \left(
      \ket{0}_0 \otimes \tensorprod_{k=1}^N \ket{0}_k
   \right)
\nonumber \\
   +
   &\sin \left(\frac{\theta}{2}\right)&
   \left(
     \cos(2\sqrt{N}t)
   \ket{1}_0 \otimes \tensorprod_{k=1}^N \ket{0}_k
   -i\sin(2\sqrt{N}t) 
   \ket{0}_0 \otimes
   \frac{1}{\sqrt{N}}\sum\limits_{l=1}^N
   \ket{0,\ldots 0,1_l,0\ldots}
   \right).
\end{eqnarray}
\end{widetext}
The rotation of the central spin indeed controls the entanglement
behavior of the system: for $\theta=0$ no entanglement is created,
while for $\theta=\pi$ the maximal entanglement oscillation will
appear. The state is a superposition of a product and an entangled
state depending on $\theta$, thus this parameter controls the
available entanglement continuously. 

\begin{figure*}[htbp]
    \centering
    \includegraphics{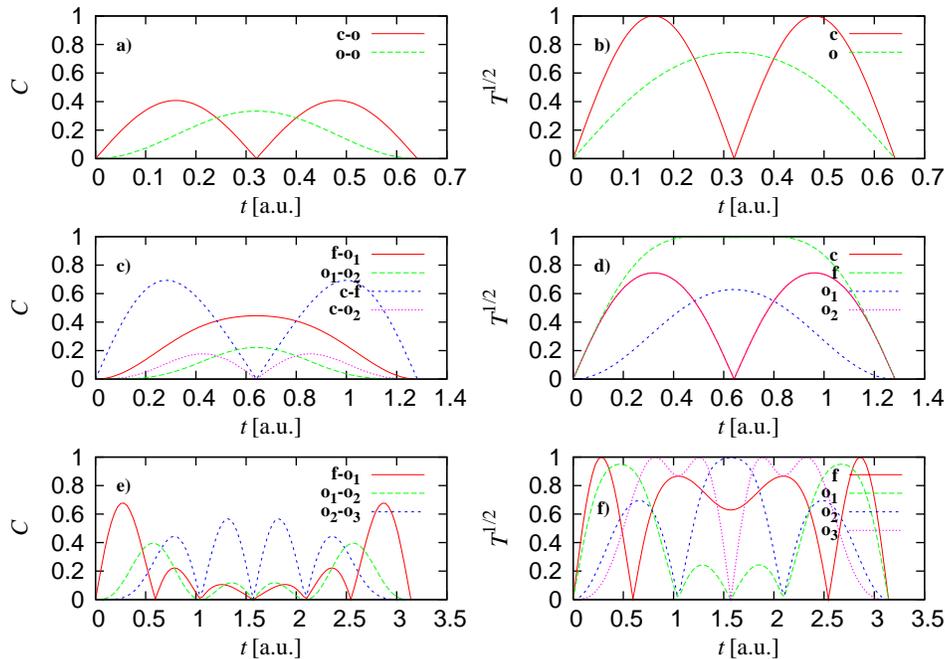}
    \caption{(Color online.)  
      Concurrence and one-tangle for spins coupled by XY
      interactions in the absence of magnetic field. In Figs. a)-d)
      6+1 spins are ordered into a star topology, while in e)-f) a
      ring of 6 spins is considered. In the initial state all spins
      are up, except for one, which is down. In a)-b) the central spin
      while in c)-d) an outer spin is flipped to point upwards.
      Figures on the left display concurrences of qubit pairs, those
      on the right display square roots of one-tangles as a function
      of time.  Legend: c: the central central spin, f: an outer spin
      which is flipped initially, o$_k$: an outer spin which is the
      $k$-th neighbor of the initially flipped one. Time is measured
      in arbitrary units, the other quantities are dimensionless. The
      figure is obtained from exact numerical diagonalization and
      direct calculations.}
    \label{fig:XYfig}
  \end{figure*}

These entanglement oscillations are different than those in case of
Ising couplings. As shown in Appendix~\ref{app:rankone}, concurrence
is equal to concurrence of assistance in the case of any superposition
of the computational basis states with all spins up and one down. This
means that in the states arising throughout this evolution
measurements do not facilitate ``focusing'' entanglement onto two
spins. Besides, it has been proven in Ref.~\cite{CoffmanKW00} that
these states saturate CKW inequalities in Eq.~\eqref{eq:CKW}, thus the
bipartite entanglement present in the states is maximal. This scheme
provides a dynamical way of preparing multipartite states with maximal
bipartite entanglement, which is controlled by the initial state of
one spin. In addition, it illustrates that Proposition~\ref{thm:ckw}
works for more than two subsystems, which is shown exactly in this
specific case.  Note that at certain times the central spin gets
disentangled from the outer ring, which is meanwhile in a state with
highest pairwise concurrence possible. Such a maximally entangled
state is reached for the whole system, too, at different times, see
also in Fig.~\ref{fig:XYfig}/a).

In Fig.~\ref{fig:XYfig} we present the behavior of concurrence and
square root of one tangles for a ring topology, and for an outer spin
in a state different from the others, as an illustration. Here we
consider the initial state producing the maximal entanglement, that
is, one spin is considered to point downwards, while all the others
point upwards. An analytical solution similar to that in
Appendix~\ref{app:andyn} would be feasible too, but more energy
eigenstates have nonzero weights in the initial state. Of course the
functions are not equal for all the spins in such case, but their
behavior is similar to the star topology. According to
Appendix~\ref{app:rankone}, concurrence is equal to concurrence of
assistance, and of course CKW inequalities are saturated.
  
  From the above discussion one might conclude that the XY couplings
  ``prefer'' to generate pure bipartite entanglement. This is however
  not the case. In order to examine this issue, we have plotted the
  behavior of entanglement quantities for an XY-coupled star
  configuration with the initial state in
  Eq.~\eqref{eq:instate_Ising}, that is, the polarized state arising
  as a product of all the spins in the same state which is a
  superposition of $\ket{0}$ and $\ket{1}$.  It appears that in this
  case concurrence between two outer spins is heavily suppressed, but
  concurrence of assistance takes rather high values for certain
  initial states. Moreover, concurrence of assistance is very close to
  the square-root of one-tangle, just as in the case of the Ising
  couplings. Thus XY couplings can, if the initial state is suitably
  chosen, produce states with a high amount of bipartite entanglement
  available via assistive measurements. Notice however, that the
  square-root of one-tangle is higher than concurrence of assistance,
  thus there is also some multipartite entanglement present in the
  system which cannot be accessed by assistive measurements.
\begin{figure*}[htbp]
  \centering
      \includegraphics{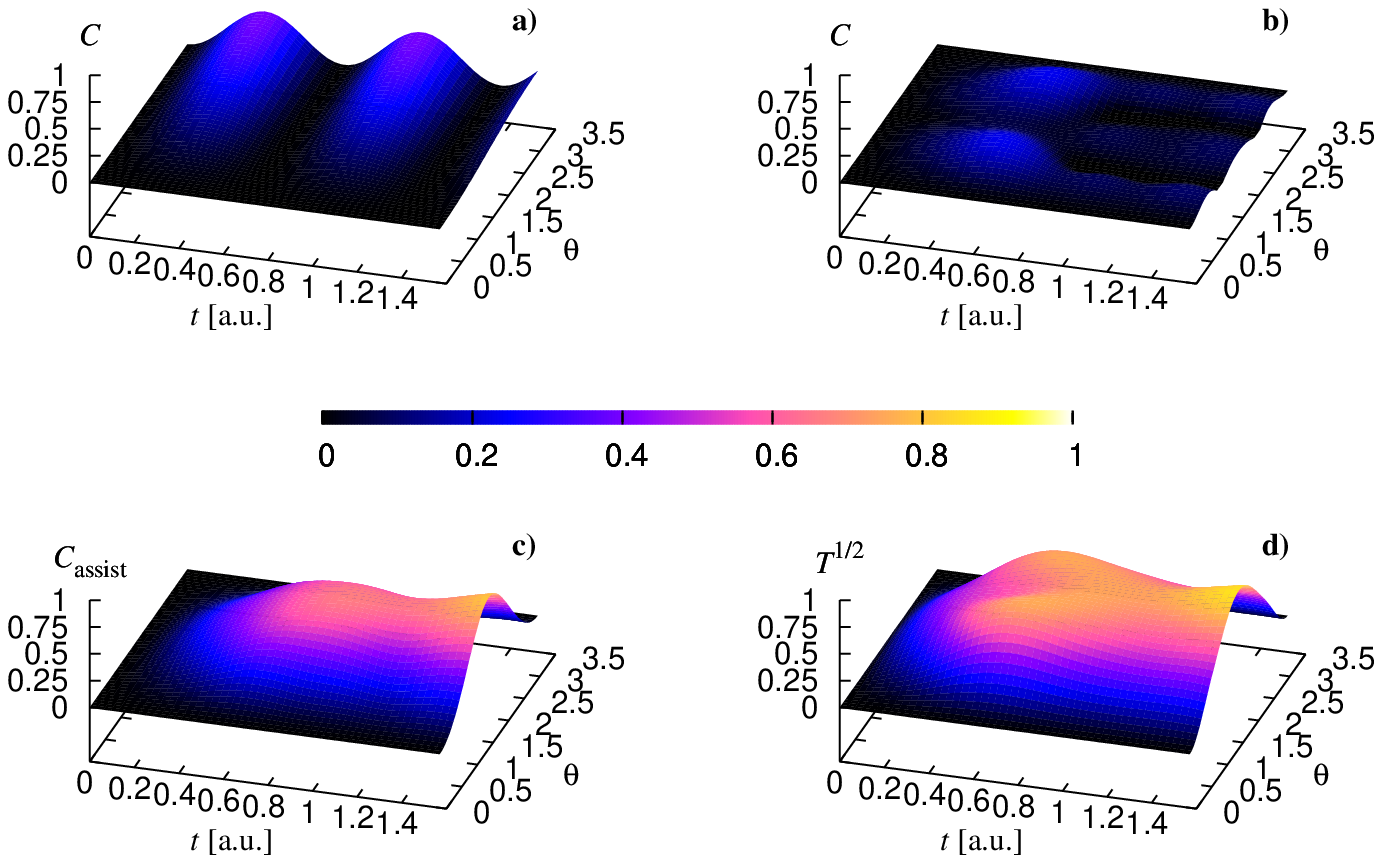}
  \caption{(Color online.) 
    Comparison of rotating all spins or the central spin in the
    initial state of a 6+1 spin star with XY couplings. Fig. a)
    displays the temporal behavior of concurrence if the central spin
    is rotated, i.e. the initial state in Eq.~\eqref{eq:inXY} is used,
    while the other three figures display the evolution of
    concurrence, concurrence of assistance and square-root of
    one-tangle with an initial state in Eq.~\eqref{eq:instate_Ising},
    that is, all spins in the same superposition of $\ket{0}$ and
    $\ket{1}$. All the bipartite quantities correspond to two outer
    spins, square-root of one-tangle is that of one of these. $\theta$
    stands for the dimensionless parameter of the input state.}
  \label{fig:xyallcontrol}
\end{figure*}

\begin{figure*}[htbp]
    \centering
      \includegraphics{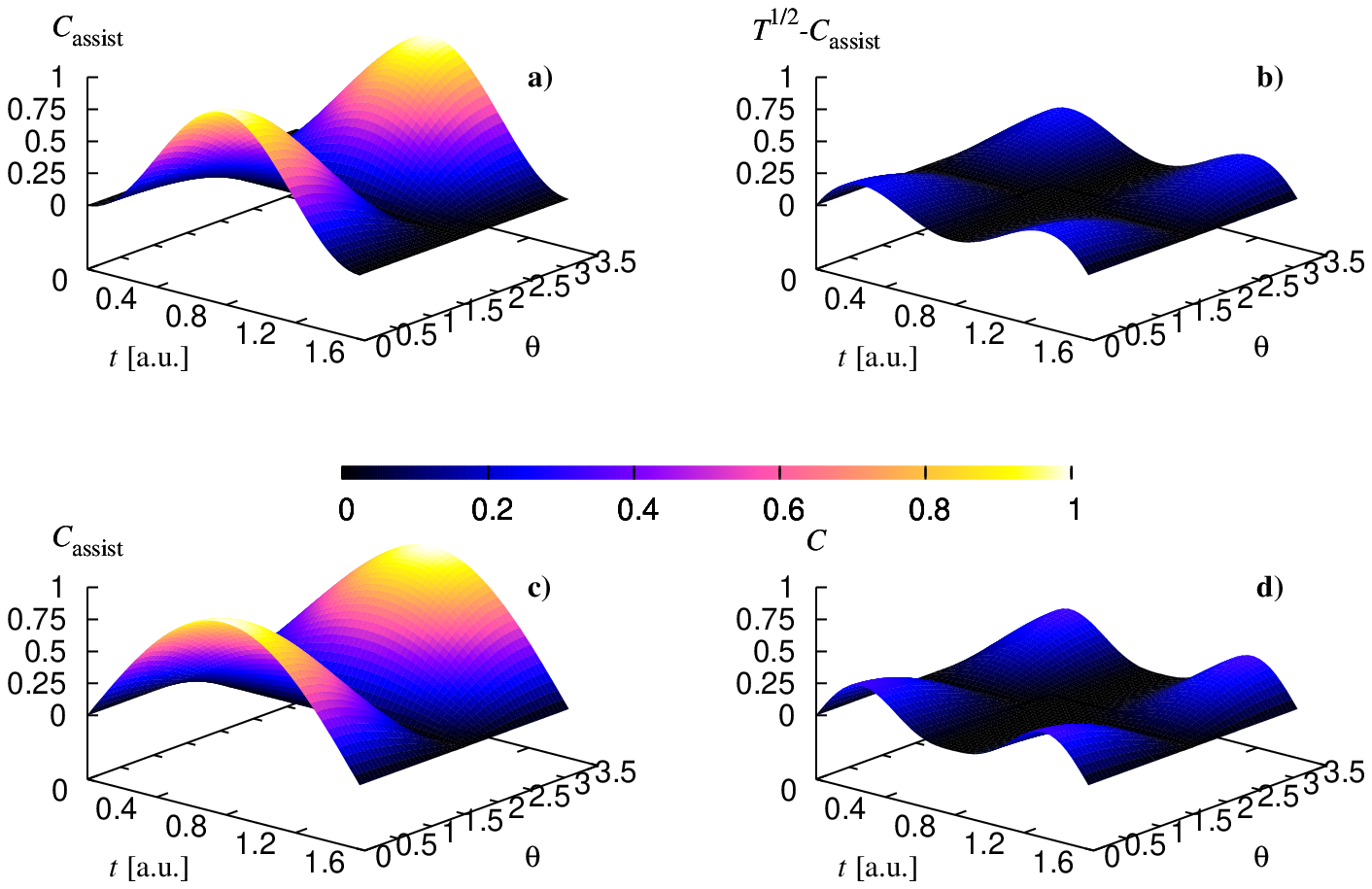}    
    \caption{(Color online.) 
      Control of entanglement generation in a system of 6+1
      Ising-coupled spins in a star configuration. The central spin is
      rotated, i.e. initial state is that in Eq.~\eqref{eq:inXY}, the
      others are in the state $\ket{0}$. Figures a) and c) display
      temporal behavior of concurrence as a function of parameter
      $\theta$ of the initial state, for a) two outer spins  and
      b) an outer and a central spin. Figure b) shows the difference
      between square root of one tangle and concurrence of assistance
      for two outer spins. Figure d) shows concurrence for the central
      and an outer spin. This quantity is zero for the outer spins.}
    \label{fig:Isingcontrol}
  \end{figure*}

Consider now Ising interactions, and ask whether it is sufficient to
rotate just one spin in order to control the amount of available
entanglement, e.g. disable entanglement oscillations.  For the
rotation of an outer spin in the star configuration or the ring
topology we have found that entanglement cannot be completely
suppressed. However, if we rotate the central spin in a star topology,
it is possible to control entanglement behavior. This is illustrated
in Fig.~\ref{fig:Isingcontrol}. Similarly to the case of initial state
of~\eqref{eq:instate_Ising}, concurrence of assistance is almost equal
to the square root of one-tangle, while concurrence itself is close to
zero.

\begin{figure}[htbp]
  \centering
  \includegraphics{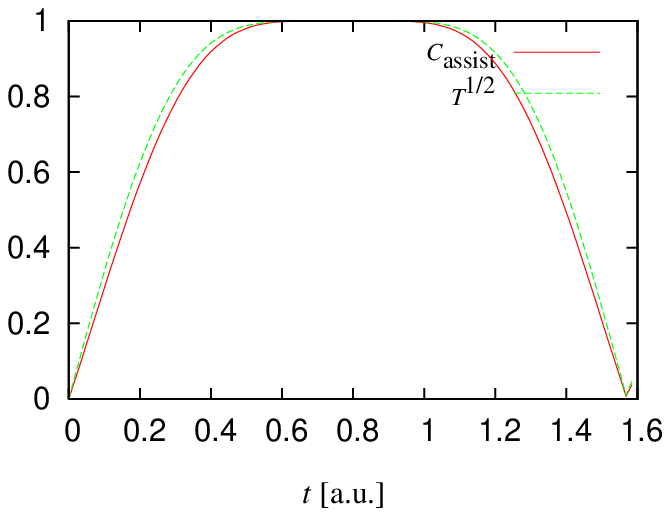}
  \caption{(Color online.) 
    Time evolution of concurrence of assistance and one-tangle for the
    ``weird'' Hamiltonian in Eq.~\eqref{eq:weird}, for 6 spins. In the
    initial product state all spins point upwards.}
  \label{fig:weird}
\end{figure}

It is important to note that the possible high value of concurrence of
assistance appears to have nothing to do with the bipartite nature of
the couplings. In order to see this, consider a ring of spins with the
``weird'' threepartite couplings
\begin{equation}
  \label{eq:weird}
  \hat H_{\text{weird}}= -\sum_k \sx_{k-1} \sy_{k} \sx_{k+1}.
\end{equation}
The temporal behavior of concurrence of assistance and square-root of
one-tangle for neighbors is shown in
Fig.~\ref{fig:weird}. Concurrence of assistance apparently reaches its
upper limit showing that threepartite interaction can also generate
maximal focusable bipartite entanglement.

In this Section we have shown that it is possible to generate
entanglement oscillations not only between product and graph (or
cluster) states, but also between product states, and states with
maximal possible bipartite entanglement, and control this entanglement
behavior by the initial state.

\section{Entangled bases in the presence of a magnetic field}
\label{sect:bases}

In Section~\ref{sect:graphstates} we have seen that in the absence of
magnetic field the Ising couplings induce such dynamics that
\emph{all} the states of the computational basis evolve into graph
states periodically. In the Heisenberg picture we may interpret this
so that the product of the $\sz$ operators evolves to such a joint
observable, which has an eigenbasis formed fully by graph states. One
of the key features of such states is that they can be projected onto
a maximally entangled state of any pair of selected spins by a von
Neumann measurement on the rest of the spins. We show here that this
property is preserved, moreover enhanced if the magnetic field is
present.

First we consider the Ising Hamiltonian with a magnetic field pointing
towards a direction characterized by the angle $\phi$:
\begin{equation}
  \label{eq:Ising}
    \hat H _\text{Ising}= -\sum\limits_{\langle k,l \rangle}
  \sx_k \otimes \sx_l -
  B\sum_k e^{i\frac{\phi}{2}\sx_k} \sz_k e^{-i\frac{\phi}{2}\sx_k}.
\end{equation}
Thus we have two free parameters characterizing the magnetic field,
its magnitude $B$ and direction $\phi$. Note that the rotation of the
magnetic field is equivalent to a rotation of the initial state in
this case. 

In particular, we are interested in the temporal behavior of the
concurrence of assistance $C_{\text{assist}}$ for certain pairs of
spins. Therefore we calculate the time evolution of all the states
$\ket{e_i}$ of the computational basis:
\begin{equation}
  \label{eq:isingtrstates}
  \Ket{e_i'(B,t)}=
  \exp\left(-i\hat H _{\text{Ising}}t\right)\Ket{e_i},
  \quad i=1\ldots 2^N,
\end{equation}
Then we can evaluate the average
\begin{equation}
  \label{eq:ensavg}
  {\overline{C_{\text{assist}}}}(B,t)= \frac{1}{2^N}
  \sum_i C_{\text{assist}}\left( \Ket{e_i'(B,t)}  \right),
\end{equation}
and also the standard deviation
\begin{equation}
  \label{eq:ensdev}
  \sigma_{C_{\text{assist}}}(B,t)= \sqrt{\overline{C_{\text{assist}}^2}-\overline{C_{\text{assist}}}^2}
\end{equation}
of concurrence of assistance over the computational basis states as
initial states.  The deviation is informative regarding the deviation
of the quantity from the average for the different initial states.

\begin{figure*}[htbp]
  \centering
  \includegraphics{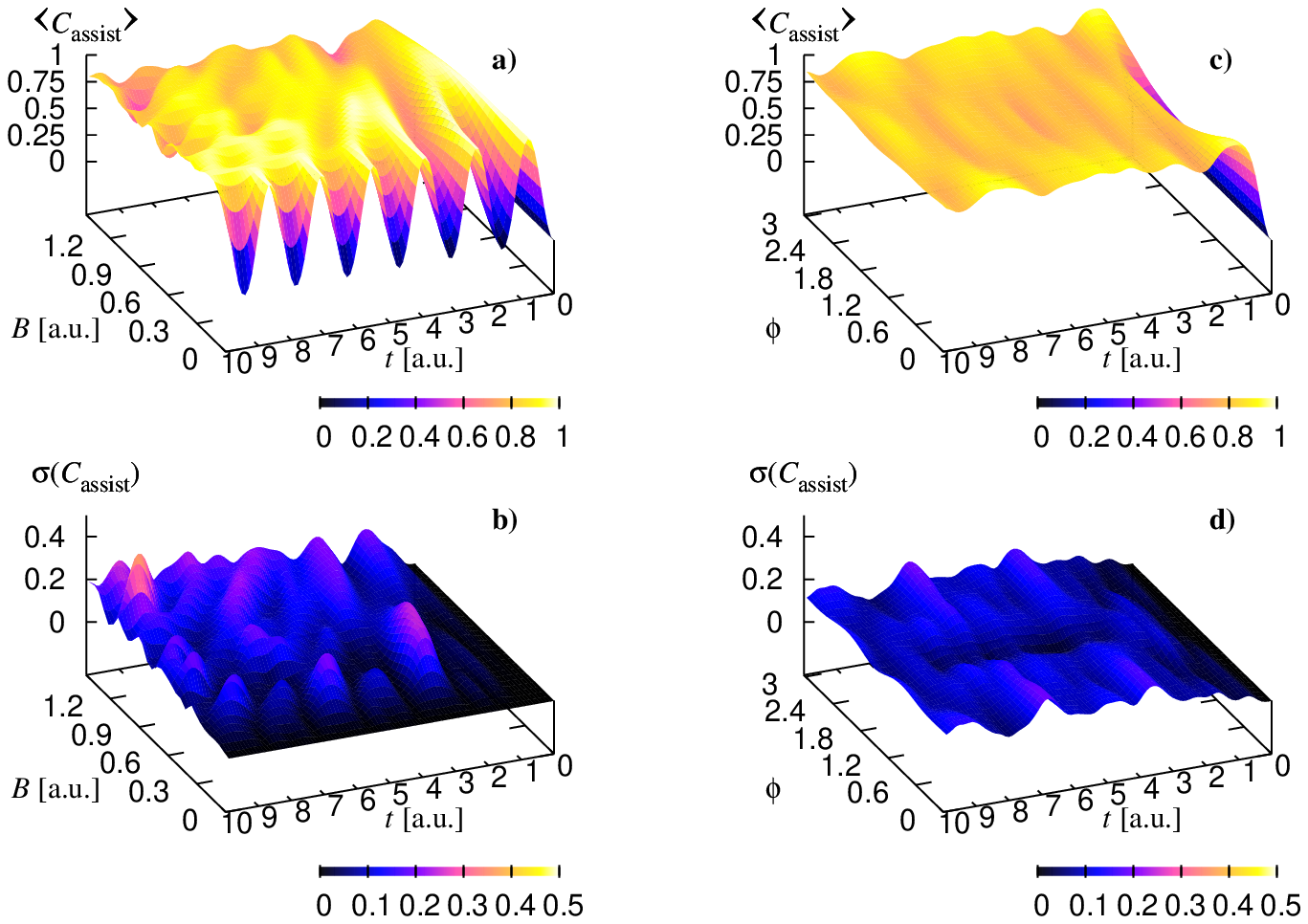}
  \caption{(Color online.) 
    Average (a,c) and standard deviation (b,d) 
    of concurrence of assistance for a pair of outer spins of a star
    topology, taken over all the possible computational basis states
    as initial states. Ising Hamiltonian with a magnetic field as in
    Eq.~\eqref{eq:Ising}, 4+1 spins in a ring topology. In Figs. a)
    and b), $\phi=0$, $B$ dependence is plotted in Figs. c) and d),
    $B=1$, $\phi$-dependence is plotted.  Similar figures are obtained
    for different choice of the spin pair, and ring topologies too.}
  \label{fig:Isingbasis}
\end{figure*}

A typical result of the calculation is plotted in
Fig.~\ref{fig:Isingbasis}
For $B=0$ the expected entanglement oscillations are present. If the
magnetic field is nonzero, the system does not tend to return to the
initial product states. Magnetic field resolves many of the the high
degeneracies of the Ising Hamiltonian, and the eigenvalues become
incommensurable. Therefore, even though the evolution of the system
will be almost periodic according to the quantum recurrence
theorem~\cite{BocchieriL57}, the reasonable approximate recurrences
occur after an extremely long time.

For $B\neq 0$, the ensemble average of concurrence of assistance
appears to be rather strictly close to one for quite long time
intervals, while its standard deviation is low. The deviation can be
further suppressed by the suitable choice of magnetic field.  This
behavior of concurrence of assistance is very similar to that in
Fig.~\ref{fig:Isingbasis} also for different chosen pair of qubits,
for qubit pairs of a ring topology, and also for different
computationally feasible number of qubits. From this we can conclude
that the elements of the computational basis are transformed into
states which can be projected into nearly maximally entangled states
of chosen two spins via a von Neumann measurement on the rest of the
spins. Otherwise speaking, Ising couplings do take the products of
$\sz$ matrices to such complete set of commuting operators, whose
eigenstates have the above mentioned property. The temporal duration
of the presence of this property is significantly enhanced by the
magnetic field.

The so arising entanglement is essentially multipartite: the
appearance of the magnetic field does not enhance concurrence of the
qubit pairs as it can be verified by performing the same calculation
with concurrence. Note that the characteristic behavior of the
entanglement as reflected by the Meyer-Wallach measure for the kicked
Ising model, also in the case of the presence of a magnetic field
pointing towards an arbitrary direction was also reported in
\cite{quantph0409039}.

Another relevant question might be whether the required measurements
are local, i.e. how much localizable entanglement is present. To
illustrate this issue in our numerical framework, we have evaluated a
lower bound for localizable entanglement by the mere consideration of
a measurement on the computational basis. According to our experience,
the behavior of the so available bipartite entanglement resembles
that of concurrence of assistance, but takes lower values. However,
quite remarkable bipartite entanglement is still available, which is
in most of the cases still higher than the limit that CKW inequalities
would allow for, without measurements.

Next we investigate the properties of the $XY$-model from the same
point of view: into Eq.~\eqref{eq:isingtrstates} we substitute the
Hamiltonian 
\begin{eqnarray}
  \label{eq:XY}
    \hat H_{\text{XY}} = -\sum\limits_{\langle k,l \rangle}
  \left( \sx_k \otimes \sx_l +\sy_k \otimes \sy_l\right) \nonumber \\
  - \sum_k e^{i\frac{\phi}{2}\sx_k} \sz_k e^{-i\frac{\phi}{2}\sx_k}.
\end{eqnarray}
A homogeneous magnetic field parallel to the $z$ does not have any
effect on the entanglement behavior of the system, as 
\begin{equation}
  \label{eq:commut}
\left[  \sum_l \sz;\sum\limits_{\langle k,l \rangle}
  \left( \sx_k \otimes \sx_l +\sy_k \otimes \sy_l\right)\right]=0
\end{equation}
thus the local rotations generated by $\sum_l \sz$ can be taken into
account after calculating the effect of the couplings.  Therefore we
pick $B=1$, and investigate the dependence of concurrence and
concurrence of assistance on the direction $\phi$ of the field.

\begin{figure*}
  \centering
  \includegraphics{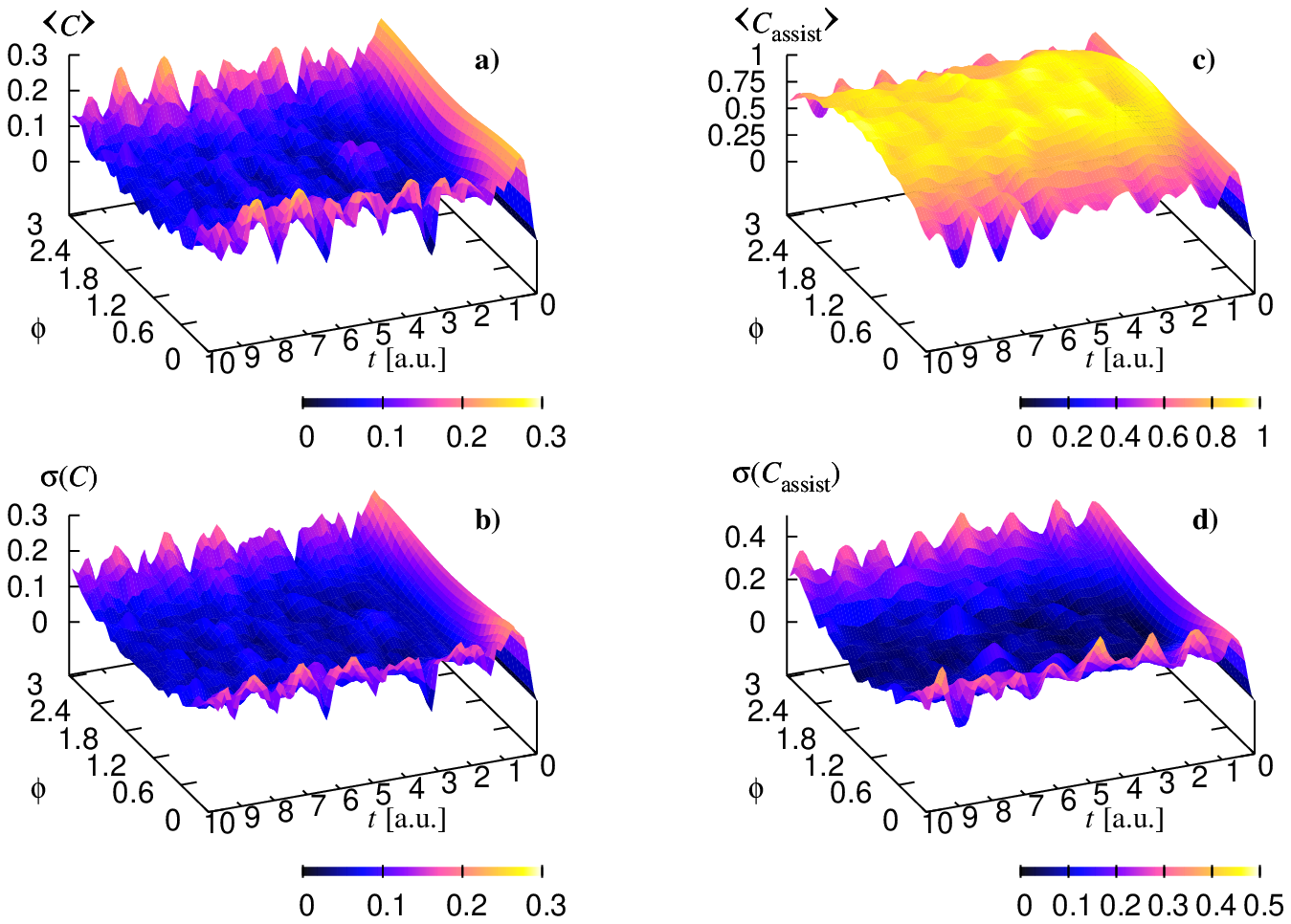}
  \caption{(Color online.) 
    Time evolution of averages (a,c) and deviations (b,d) 
    of concurrence (a,b) and concurrence of assistance (c,d) for two
    outer spins of a star configuration of 4+1 spins coupled by the XY
    Hamiltonian with magnetic field in \eqref{eq:XY}. Parameter $\phi$
    describes the direction of the magnetic field. Similar behavior
    was observed for ring topologies and different choice of the qubit
    pair too.}
  \label{fig:XYbasis}
\end{figure*}

The quantities evaluated are again those in Eqs.~\eqref{eq:ensavg}
and~\eqref{eq:ensdev}, both for concurrence and concurrence of
assistance. A typical result is displayed in Fig.~\ref{fig:XYbasis}.
It appears that for $\phi=0$ we obtain oscillations in the average
concurrence, too, while concurrence of assistance is not significantly
higher than concurrence itself. The appropriate choice of the
direction of the magnetic field can suppress concurrence,
significantly enhance concurrence of assistance and decrease its
deviation. Thus even though the couplings are not Ising type, at least
the feature of the Ising couplings that it produces bases with high
concurrence of assistance can be retained.

\section{Conclusions}
\label{sect:concl}

In this paper we have related the problems of maximizing pairwise
concurrence and pairwise concurrence of assistance in a system of
multiple qubits.  We have shown that the square root of one tangle of
a qubit is an upper bound for the concurrence of assistance of a qubit
pair containing the particular qubit. We have also shown that for a
certain set of states for which the CKW inequality is known to be
saturated, the concurrence is equal to the concurrence of assistance.
This means that the bipartite subsystem under consideration is not
correlated with the rest of the system via intrinsic multipartite
entanglement.

We have also studied the entanglement behavior of spin-1/2 systems
modeling qubits, from this perspective. We have shown that in a star
configuration of an XY coupled spins entanglement oscillations between
product states and states with maximal bipartite entanglement
according to CKW inequalities can be dynamically generated. The
oscillations can be controlled by rotating the spin which mediates the
interaction, and at some points it gets disentangled from the rest of
the outer ring, which is maximally entangled in the CKW sense. This
maximal entanglement is reached for the whole system, too.  We have
shown numerically that the star topology facilitates the similar
control of entanglement oscillations between product and graph states.
The rotation of all the qubits of the initial state on the other hand
leads to different behavior of concurrence of assistance, as the
enhancement of bipartite entanglement to the measurement appears. We
have found similar behavior for different topologies numerically.

According to our numerical results magnetic field can lead to the
temporal enhancement of concurrence of assistance in the entanglement
oscillations starting from the states of the computational basis, in
the case of spins coupled by Ising interactions, arranged into ring or
star topologies.  Thereby a special entangled basis can be accessed.
We have found similar behavior for the case of XY couplings: magnetic
field applied along properly chosen direction suppresses concurrence
and enhances concurrence of assistance.

According to the presented results, pairwise couplings between spins
and qubits can be used effectively for different tasks of distributing
bipartite entanglement between multiple parties. It is also possible
to control the dynamical behavior of entanglement by local quantum
operations such as rotation of control qubits. Besides, magnetic
field can be utilized to temporally enhance certain entanglement
features, or to chose between qualitatively different kinds of
entanglement behavior. It would be also interesting to investigate
whether the entangled bases available in the described means are
useful for quantum information processing tasks.

\begin{acknowledgments}
This work was supported by the European Union projects QGATES and
CONQUEST, and by the Slovak Academy of Sciences via the project CE-PI.
M.~K. acknowledges the support of National Scientific Research Fund of
Hungary (OTKA) under contracts Nos. T043287 and T034484. The authors
thank G\'eza T\'oth for useful discussions.  
\end{acknowledgments}

\appendix  

\section{An inequality}
\label{app:ineqproof}

In this Appendix we show, that for two Hermitian, positive
semidefinite $2\times 2$ matrices $\hat A$ and $\hat B$,
\begin{equation}
  \label{eq:mainineq}
  \sqrt{\det \hat A} +   \sqrt{\det \hat B}
  \leq
  \sqrt{\det\left( \hat A + \hat B \right)}
\end{equation}
holds.

First we remark that for the square root of a Hermitian positive
semidefinite matrix
\begin{equation}
  \det \sqrt{\hat A} = \sqrt{\det \hat A}.
\end{equation}
Thus we can rewrite inequality~\eqref{eq:mainineq} as
\begin{equation}
  \det \sqrt{\hat A} +   \det \sqrt{\hat B}
  \leq
  \sqrt{\det\left( \hat A + \hat B \right)},
\end{equation}
or equivalently,
\begin{equation}
\label{eq:prineq1}
  \left(\det \sqrt{\hat A} +   \det \sqrt{\hat B}\right)^2
  \leq
  \det\left( \hat A + \hat B \right).
\end{equation}
Without the loss of generality we can perform calculations on the
eigenbasis of $\hat A$. Thus we can introduce the notation
\begin{eqnarray}
  \sqrt{\hat A}=a
  \begin{pmatrix}
    a_1 & 0 \cr 0 & 1-a_1
  \end{pmatrix}
  \nonumber \\
  \sqrt{\hat B}=b
  \begin{pmatrix}
    b_1 & b_2 \cr b_2^* & 1-b_1
  \end{pmatrix}
\label{eq:matrnot},
\end{eqnarray}
where $a$, $b$, $a_1$ and $b_1$ are real. Substituting Eq.~\eqref{eq:matrnot}
into Eq.~\eqref{eq:prineq1}, after some calculation we obtain
\begin{eqnarray}
  \det\left( \hat A + \hat B \right)-  \left(\det \sqrt{\hat A} -   \det \sqrt{\hat B}\right)^2 \nonumber \\
= a^2b^2\left(|b_2|^2+(a_1-b_1)^2\right) \geq 0,
\end{eqnarray}
which is always justified. 

Note that the inequality just proven is a special property of $2\times
2$ matrices: if we replaced $\hat A$ and $\hat B$ by positive
numbers as ```$1\times 1$'' matrices, the direction of
inequality~\eqref{eq:mainineq} would be reverse.

\section{Analytical solution for the $XY$-coupled star} 
\label{app:andyn} 

Here we derive the time evolution for our specific input states in an
XY coupled star configuration, based on
Refs.~~\cite{HuttonB04,BreuerBP04}. Consider the Hamiltonian in
Eq.~\eqref{eq:XYnoB} for a star topology of $N+1$ spins. Let spin 0 be
the central one, thus the Hamiltonian reads
\begin{equation}
  \hat H_{XY}=-\sx_0\sum_{k=1}^N \sx_k  -  \sy_0 \sum_{k=1}^N \sy_k.
\end{equation}
Introducing the joint spin operators of the outer spins
\begin{eqnarray}
  \label{eq:jouter}
  \hat J_{\alpha}=\sum_{k=1}^N 
  \frac12 \hat \sigma _\alpha , \qquad \alpha=x,y,z,
  \nonumber \\
  \hat J_{\pm}=\sum_{k=1}^N \hat J_x \pm i \hat J_y
\end{eqnarray}
and the operator for the $z$ component of the total angular momentum 
\begin{equation}
  \label{eq:totang}
  \hat L_z = \hat J_z + \frac12  \sz,
\end{equation}
the following commutation relations hold:
\begin{equation}
  \label{eq:xycomm}
  \left[ \hat H_{XY},\hat L_z \right] =
  \left[ \hat H_{XY},\hat J^2 \right] =0.
\end{equation}
Therefore the computational basis states with equal spins down span
invariant subspaces of the evolution, and the outer spins behave
collectively as one big spin. It is convenient to rewrite the
Hamiltonian in the Jaynes-Cummings type form
\begin{equation}
  \label{eq:Hrev}
  \hat H_{XY}=-\left(
    \Sp_0\hat J_- + \Sm_0 \hat J_+,
\right)
\end{equation}
which has the eigenvalues and eigenvectors
\begin{eqnarray}
  \label{eq:Heig}
  \ket{\phi_{j,m,\pm}}=\frac{1}{\sqrt2}\left(
    \ket{1}\ket{j;m} \pm  \ket{0}\ket{j;m-1}
\right) \nonumber \\
 \omega_{j,m,\pm}=\mp 2\sqrt{(j+m)(j-m+1)},
\end{eqnarray}
where the states $\ket{j;m}$ are the eigenvectors for the outer spins,
while the states $\ket{0}$ and $\ket{1}$ are the states of the central
spin in our qubit notation.  (Note that in our notation,
$\ket{0}=\ket{\uparrow}$, thus $\Sp \ket{1}=2\ket{0}$.)

We consider the possibility of the control by the rotation of the
central spin, thus our initial state reads
\begin{widetext}
\begin{equation}
  \label{eq:inXYalt}
  \Ket{\Psi(t=0)}=
  A_0\Ket{0} \otimes \Ket {00\ldots} 
  + A_1\Ket{1} \otimes \Ket {00\ldots}
  =
  A_0\Ket{0} \otimes \Ket {j=\frac{N}{2},m=\frac{N}{2}} 
  + A_1\Ket{1} \otimes \Ket {j=\frac{N}{2},m=\frac{N}{2}}.
\end{equation}
Rewriting this in the energy basis in Eq.~\eqref{eq:Heig} we obtain
\begin{eqnarray}
\label{eq:enbas}
  \Ket{\Psi(t=0)}=
  A_0\Ket{0} \otimes \Ket {j=\frac{N}{2},m=\frac{N}{2}} 
  + \frac{A_1}{2}\left(
 \Ket{1} \otimes \Ket {j=\frac{N}{2},m=\frac{N}{2}} + 
 \Ket{1} \otimes \Ket {j=\frac{N}{2},m=\frac{N}{2}-1}\right)&  \nonumber \\
  + \frac{A_1}{2}\left(
 \Ket{1} \otimes \Ket {j=\frac{N}{2},m=\frac{N}{2}} -
 \Ket{1} \otimes \Ket {j=\frac{N}{2},m=\frac{N}{2}-1}\right).&
\end{eqnarray}
Substituting the $e^{-i\omega t}$ factors, where according to
Eq.~\eqref{eq:Heig}, the $\omega$-s are $0,-2\sqrt{N},+2\sqrt{N}$ for
the three summands of Eq.~\eqref{eq:enbas} respectively, after some
algebra we obtain
\begin{equation}
  \label{eq:appevol}
  \Ket{\Psi(t=0)}=
  A_0\Ket{0} \otimes \Ket{0,0,\ldots} 
  + A_1 \left( \cos(2\sqrt{N}t) \ket{1}\otimes\ket{0,0,\ldots}
                      -i\sin (2\sqrt{N}t) \ket{0} \otimes 
                      \Ket{j=\frac{N}{2}, m=\frac{N}{2}-1}\right).
\end{equation}
This shows that the complex phases of $A_0$ and $A_1$ are irrelevant
from the point of view of the entanglement properties. Substituting
$A_0=\cos(\theta/2)$ and $A_1=\sin(\theta/2)$ into
Eq.~\eqref{eq:appevol}, and calculating $\ket{j=N/2,m=N/2-1}$ by
applying $\hat J_-$ on $\ket{j=N/2,m=N/2}$, we obtain
Eq.~\eqref{eq:XYtime}, the desired result.
\end{widetext}

\section{Relation of concurrence and concurrence of assistance for
  states with maximum one spin down}
\label{app:rankone}

In this appendix we show that for states of $N$ qubits of the form
\begin{equation}
  \label{eq:oneup}
  \Ket{\Psi_1}=\sum_{k=0}^N A_k \ket{\underline{k}}
\end{equation}
where 
\begin{eqnarray}
  \ket{\underline{0}}&=&\ket{0,0,\ldots, 0},\nonumber \\
  \ket{\underline{k}}&=&\ket{0,\ldots 0,1_k,0\ldots}, \quad k\neq 0,
\end{eqnarray}
concurrence equals to concurrence of assistance for any pairs of the
qubits. 

Consider the spins $k$ and $l$. Their density matrix in the
computational basis is of the form
\begin{equation}
  \varrho^{(kl)}=
  \begin{pmatrix}
    \varrho_{00,00}   & \varrho_{00,01}   & \varrho_{00,10} & 0  \cr
    \varrho_{00,01}^* & \varrho_{01,01}   & \varrho_{01,10} & 0  \cr
    \varrho_{00,10}^* & \varrho_{01,10}^* & \varrho_{10,10} & 0  \cr
            0         &       0           &      0          & 0  
  \end{pmatrix}
\end{equation}
Direct calculation of concurrence and concurrence of assistance
according to Eqs.~\eqref{eq:concurrence} and Eq.~\eqref{eq:cassist}
yields
\begin{equation}
 \label{eq:b1}
  C_{kl}=2|\varrho_{01,10}|, \qquad 
  C^{\text{assist}}_{kl}= 2\sqrt{\varrho_{01,01}\varrho_{10,10}}.
\end{equation}
Calculating the required matrix elements from Eq.~\eqref{eq:oneup} we
find 
\begin{equation}
  \label{eq:b2}
  \varrho_{01,01}=A_k^*A_k, \quad
  \varrho_{10,10}=A_l^*A_l, \quad
  \varrho_{01,10}=A_k^*A_l.
\end{equation}
Substituting Eq.~\eqref{eq:b2} into Eq.~\eqref{eq:b1} gives $C_{kl}=
C^{\text{assist}}_{kl}$ for arbitrary k,l.


\begin{thebibliography}{37}
\expandafter\ifx\csname natexlab\endcsname\relax\def\natexlab#1{#1}\fi
\expandafter\ifx\csname bibnamefont\endcsname\relax
  \def\bibnamefont#1{#1}\fi
\expandafter\ifx\csname bibfnamefont\endcsname\relax
  \def\bibfnamefont#1{#1}\fi
\expandafter\ifx\csname citenamefont\endcsname\relax
  \def\citenamefont#1{#1}\fi
\expandafter\ifx\csname url\endcsname\relax
  \def\url#1{\texttt{#1}}\fi
\expandafter\ifx\csname urlprefix\endcsname\relax\def\urlprefix{URL }\fi
\providecommand{\bibinfo}[2]{#2}
\providecommand{\eprint}[2][]{\url{#2}}

\bibitem[{\citenamefont{Koashi et~al.}(2000)\citenamefont{Koashi, Bu{\v z}ek,
  and Imoto}}]{KoashiBI00}
\bibinfo{author}{\bibfnamefont{M.}~\bibnamefont{Koashi}},
  \bibinfo{author}{\bibfnamefont{V.}~\bibnamefont{Bu{\v z}ek}},
  \bibnamefont{and} \bibinfo{author}{\bibfnamefont{N.}~\bibnamefont{Imoto}},
  \bibinfo{journal}{Phys. Rev. A} \textbf{\bibinfo{volume}{62}},
  \bibinfo{pages}{050302} (\bibinfo{year}{2000}).

\bibitem[{\citenamefont{Plesch and Bu{\v z}ek}(2003)}]{PleschB03}
\bibinfo{author}{\bibfnamefont{M.}~\bibnamefont{Plesch}} \bibnamefont{and}
  \bibinfo{author}{\bibfnamefont{V.}~\bibnamefont{Bu{\v z}ek}},
  \bibinfo{journal}{Phys. Rev. A} \textbf{\bibinfo{volume}{67}},
  \bibinfo{pages}{012322} (\bibinfo{year}{2003}).

\bibitem[{\citenamefont{Plesch and Bu{\v z}ek}(2002)}]{PleschB02}
\bibinfo{author}{\bibfnamefont{M.}~\bibnamefont{Plesch}} \bibnamefont{and}
  \bibinfo{author}{\bibfnamefont{V.}~\bibnamefont{Bu{\v z}ek}},
  \bibinfo{journal}{Quantum Inform. Comput.} \textbf{\bibinfo{volume}{2}},
  \bibinfo{pages}{530} (\bibinfo{year}{2002}).

\bibitem[{\citenamefont{Coffman et~al.}(2000)\citenamefont{Coffman, Kundu, and
  Wootters}}]{CoffmanKW00}
\bibinfo{author}{\bibfnamefont{V.}~\bibnamefont{Coffman}},
  \bibinfo{author}{\bibfnamefont{J.}~\bibnamefont{Kundu}}, \bibnamefont{and}
  \bibinfo{author}{\bibfnamefont{W.~K.} \bibnamefont{Wootters}},
  \bibinfo{journal}{Phys. Rev. A} \textbf{\bibinfo{volume}{61}},
  \bibinfo{pages}{052306} (\bibinfo{year}{2000}).

\bibitem[{\citenamefont{O'Connor and Wootters}(2001)}]{OConnorW01}
\bibinfo{author}{\bibfnamefont{K.~M.} \bibnamefont{O'Connor}} \bibnamefont{and}
  \bibinfo{author}{\bibfnamefont{W.~K.} \bibnamefont{Wootters}},
  \bibinfo{journal}{Phys. Rev. A} \textbf{\bibinfo{volume}{63}},
  \bibinfo{pages}{052302} (\bibinfo{year}{2001}).

\bibitem[{\citenamefont{Briegel and Raussendorf}(2001)}]{BriegelR01}
\bibinfo{author}{\bibfnamefont{H.~J.} \bibnamefont{Briegel}} \bibnamefont{and}
  \bibinfo{author}{\bibfnamefont{R.}~\bibnamefont{Raussendorf}},
  \bibinfo{journal}{Phys. Rev. Lett.} \textbf{\bibinfo{volume}{86}},
  \bibinfo{pages}{910} (\bibinfo{year}{2001}).

\bibitem[{\citenamefont{Hein et~al.}(2004)\citenamefont{Hein, Eisert, and
  Briegel}}]{HeinEB04}
\bibinfo{author}{\bibfnamefont{M.}~\bibnamefont{Hein}},
  \bibinfo{author}{\bibfnamefont{J.}~\bibnamefont{Eisert}}, \bibnamefont{and}
  \bibinfo{author}{\bibfnamefont{H.~J.} \bibnamefont{Briegel}},
  \bibinfo{journal}{Phys. Rev. A} \textbf{\bibinfo{volume}{69}},
  \bibinfo{pages}{062311} (\bibinfo{year}{2004}).

\bibitem[{\citenamefont{Raussendorf and Briegel}(2001)}]{RaussendorfB01}
\bibinfo{author}{\bibfnamefont{R.}~\bibnamefont{Raussendorf}} \bibnamefont{and}
  \bibinfo{author}{\bibfnamefont{H.~J.} \bibnamefont{Briegel}},
  \bibinfo{journal}{Phys. Rev. Lett.} \textbf{\bibinfo{volume}{86}},
  \bibinfo{pages}{5188} (\bibinfo{year}{2001}).

\bibitem[{\citenamefont{Raussendorf et~al.}(2003)\citenamefont{Raussendorf,
  Browne, and Briegel}}]{RaussendorfBB03}
\bibinfo{author}{\bibfnamefont{R.}~\bibnamefont{Raussendorf}},
  \bibinfo{author}{\bibfnamefont{D.~E.} \bibnamefont{Browne}},
  \bibnamefont{and} \bibinfo{author}{\bibfnamefont{H.~J.}
  \bibnamefont{Briegel}}, \bibinfo{journal}{Phys. Rev. A}
  \textbf{\bibinfo{volume}{68}}, \bibinfo{pages}{022312}
  (\bibinfo{year}{2003}).

\bibitem[{\citenamefont{DiVincenzo et~al.}(1998)\citenamefont{DiVincenzo,
  Fuchs, Mabuchi, Smolin, Thapliyal, and Uhlmann}}]{DiVincensoFMSTU}
\bibinfo{author}{\bibfnamefont{D.~P.} \bibnamefont{DiVincenzo}},
  \bibinfo{author}{\bibfnamefont{C.~A.} \bibnamefont{Fuchs}},
  \bibinfo{author}{\bibfnamefont{H.}~\bibnamefont{Mabuchi}},
  \bibinfo{author}{\bibfnamefont{J.~A.} \bibnamefont{Smolin}},
  \bibinfo{author}{\bibfnamefont{A.}~\bibnamefont{Thapliyal}},
  \bibnamefont{and} \bibinfo{author}{\bibfnamefont{A.}~\bibnamefont{Uhlmann}},
 (\bibinfo{publisher}{Springer-Verlag},
  \bibinfo{year}{1998}), vol. \bibinfo{volume}{1509} of
  \emph{\bibinfo{series}{Lecture notes in computer science}},
  \bibinfo{note}{e-print quant-ph/9803033}.

\bibitem[{\citenamefont{Laustsen et~al.}(2003)\citenamefont{Laustsen,
  Verstraete, and Van~Enk}}]{LaustsenVV03}
\bibinfo{author}{\bibfnamefont{T.}~\bibnamefont{Laustsen}},
  \bibinfo{author}{\bibfnamefont{F.}~\bibnamefont{Verstraete}},
  \bibnamefont{and} \bibinfo{author}{\bibfnamefont{S.~J.}
  \bibnamefont{Van~Enk}}, \bibinfo{journal}{Quantum Inform. Comput.}
  \textbf{\bibinfo{volume}{3}}, \bibinfo{pages}{64} (\bibinfo{year}{2003}).

\bibitem[{\citenamefont{Verstraete et~al.}(2004)\citenamefont{Verstraete, Popp,
  and Cirac}}]{VerstraetePC04b}
\bibinfo{author}{\bibfnamefont{F.}~\bibnamefont{Verstraete}},
  \bibinfo{author}{\bibfnamefont{M.}~\bibnamefont{Popp}}, \bibnamefont{and}
  \bibinfo{author}{\bibfnamefont{J.~I.} \bibnamefont{Cirac}},
  \bibinfo{journal}{Phys. Rev. Lett.} \textbf{\bibinfo{volume}{92}},
  \bibinfo{pages}{027901} (\bibinfo{year}{2004}).

\bibitem[{\citenamefont{Popp et~al.}(2004)\citenamefont{Popp, Verstraete,
  Martin-Delgado, and Cirac}}]{quantph0411123}
\bibinfo{author}{\bibfnamefont{M.}~\bibnamefont{Popp}},
  \bibinfo{author}{\bibfnamefont{F.}~\bibnamefont{Verstraete}},
  \bibinfo{author}{\bibfnamefont{M.~A.} \bibnamefont{Martin-Delgado}},
  \bibnamefont{and} \bibinfo{author}{\bibfnamefont{J.~I.} \bibnamefont{Cirac}},
  \bibinfo{note}{e-print quant-ph/0411123} (\bibinfo{year}{2004}),.

\bibitem[{\citenamefont{Bose}(2003)}]{Bose03}
\bibinfo{author}{\bibfnamefont{S.}~\bibnamefont{Bose}}, \bibinfo{journal}{Phys.
  Rev. Lett.} \textbf{\bibinfo{volume}{91}}, \bibinfo{pages}{207901}
  (\bibinfo{year}{2003}).

\bibitem[{\citenamefont{Christandl et~al.}(2004)\citenamefont{Christandl,
  Datta, Ekert, and Landahl}}]{ChristandlDEL04}
\bibinfo{author}{\bibfnamefont{M.}~\bibnamefont{Christandl}},
  \bibinfo{author}{\bibfnamefont{N.}~\bibnamefont{Datta}},
  \bibinfo{author}{\bibfnamefont{A.}~\bibnamefont{Ekert}}, \bibnamefont{and}
  \bibinfo{author}{\bibfnamefont{A.~J.} \bibnamefont{Landahl}},
  \bibinfo{journal}{Phys. Rev. Lett.} \textbf{\bibinfo{volume}{92}},
  \bibinfo{pages}{187902} (\bibinfo{year}{2004}).

\bibitem[{\citenamefont{Osborne and Nielsen}(2004)}]{OsborneN04}
\bibinfo{author}{\bibfnamefont{T.~J.} \bibnamefont{Osborne}} \bibnamefont{and}
  \bibinfo{author}{\bibfnamefont{N.}~\bibnamefont{Nielsen}},
  \bibinfo{journal}{Phys. Rev. A} \textbf{\bibinfo{volume}{69}},
  \bibinfo{pages}{052315} (\bibinfo{year}{2004}).

\bibitem[{\citenamefont{Schuch and Siewert}(2003)}]{SchuchS03}
\bibinfo{author}{\bibfnamefont{N.}~\bibnamefont{Schuch}} \bibnamefont{and}
  \bibinfo{author}{\bibfnamefont{J.}~\bibnamefont{Siewert}},
  \bibinfo{journal}{Phys. Rev. A} \textbf{\bibinfo{volume}{67}},
  \bibinfo{pages}{032301} (\bibinfo{year}{2003}).

\bibitem[{\citenamefont{Yung et~al.}(2004)\citenamefont{Yung, Leung, and
  Bose}}]{YungLB04}
\bibinfo{author}{\bibfnamefont{M.~H.} \bibnamefont{Yung}},
  \bibinfo{author}{\bibfnamefont{D.~W.} \bibnamefont{Leung}}, \bibnamefont{and}
  \bibinfo{author}{\bibfnamefont{S.}~\bibnamefont{Bose}},
  \bibinfo{journal}{Quantum Inform. Comput.} \textbf{\bibinfo{volume}{4}},
  \bibinfo{pages}{174} (\bibinfo{year}{2004}).

\bibitem[{\citenamefont{Chiara et~al.}(2004)\citenamefont{Chiara, Fazio,
  Machiavello, Montagero, and Palma}}]{ChiaraFMMM04}
\bibinfo{author}{\bibfnamefont{G.~D.} \bibnamefont{Chiara}},
  \bibinfo{author}{\bibfnamefont{R.}~\bibnamefont{Fazio}},
  \bibinfo{author}{\bibfnamefont{C.}~\bibnamefont{Machiavello}},
  \bibinfo{author}{\bibfnamefont{S.}~\bibnamefont{Montagero}},
  \bibnamefont{and} \bibinfo{author}{\bibfnamefont{G.~M.} \bibnamefont{Palma}},
  \bibinfo{journal}{Phys. Rev. A} \textbf{\bibinfo{volume}{70}},
  \bibinfo{pages}{062308} (\bibinfo{year}{2004}).

\bibitem[{\citenamefont{Garcia-Ripoll and Cirac}(2003)}]{Garcia-RipollC03}
\bibinfo{author}{\bibfnamefont{J.~J.} \bibnamefont{Garcia-Ripoll}}
  \bibnamefont{and} \bibinfo{author}{\bibfnamefont{J.~I.} \bibnamefont{Cirac}},
  \bibinfo{journal}{New J. Phys.} \textbf{\bibinfo{volume}{5}},
  \bibinfo{pages}{76} (\bibinfo{year}{2003}).

\bibitem[{\citenamefont{Amico et~al.}(2004)\citenamefont{Amico, Osterloh,
  Plastina, Fazio, and Palma}}]{AmicoOPRP04}
\bibinfo{author}{\bibfnamefont{L.}~\bibnamefont{Amico}},
  \bibinfo{author}{\bibfnamefont{A.}~\bibnamefont{Osterloh}},
  \bibinfo{author}{\bibfnamefont{F.}~\bibnamefont{Plastina}},
  \bibinfo{author}{\bibfnamefont{R.}~\bibnamefont{Fazio}}, \bibnamefont{and}
  \bibinfo{author}{\bibfnamefont{G.~M.} \bibnamefont{Palma}},
  \bibinfo{journal}{Phys. Rev. A} \textbf{\bibinfo{volume}{69}},
  \bibinfo{pages}{022304} (\bibinfo{year}{2004}).

\bibitem[{\citenamefont{Subrahmanyam}(2004)}]{Subrahmanyam04a}
\bibinfo{author}{\bibfnamefont{V.}~\bibnamefont{Subrahmanyam}},
  \bibinfo{journal}{Phys. Rev. A} \textbf{\bibinfo{volume}{69}},
  \bibinfo{pages}{034304} (\bibinfo{year}{2004}).

\bibitem[{\citenamefont{Plastina et~al.}(2004)\citenamefont{Plastina, Amico,
  Osterloh, and Fazio}}]{PlastinaAOF04}
\bibinfo{author}{\bibfnamefont{F.}~\bibnamefont{Plastina}},
  \bibinfo{author}{\bibfnamefont{L.}~\bibnamefont{Amico}},
  \bibinfo{author}{\bibfnamefont{A.}~\bibnamefont{Osterloh}}, \bibnamefont{and}
  \bibinfo{author}{\bibfnamefont{R.}~\bibnamefont{Fazio}},
  \bibinfo{journal}{New J. Phys.} \textbf{\bibinfo{volume}{6}},
  \bibinfo{pages}{124} (\bibinfo{year}{2004}).

\bibitem[{\citenamefont{Lakshminarayan and
  Subrahmanyam}(2004)}]{quantph0409039}
\bibinfo{author}{\bibfnamefont{A.}~\bibnamefont{Lakshminarayan}}
  \bibnamefont{and}
  \bibinfo{author}{\bibfnamefont{V.}~\bibnamefont{Subrahmanyam}},
  \bibinfo{note}{e-print quant-ph/0409039} (\bibinfo{year}{2004}).

\bibitem[{\citenamefont{Subrahmanyam and
  Lakshminarayan}(2004)}]{quantph0409048}
\bibinfo{author}{\bibfnamefont{V.}~\bibnamefont{Subrahmanyam}}
  \bibnamefont{and}
  \bibinfo{author}{\bibfnamefont{A.}~\bibnamefont{Lakshminarayan}},
  \bibinfo{note}{e-print quant-ph/0409048} (\bibinfo{year}{2004}).

\bibitem[{\citenamefont{Vidal et~al.}(2004)\citenamefont{Vidal, Palacios, and
  Aslangul}}]{VidalPA04}
\bibinfo{author}{\bibfnamefont{J.}~\bibnamefont{Vidal}},
  \bibinfo{author}{\bibfnamefont{G.}~\bibnamefont{Palacios}}, \bibnamefont{and}
  \bibinfo{author}{\bibfnamefont{C.}~\bibnamefont{Aslangul}},
  \bibinfo{journal}{Phys. Rev. A} \textbf{\bibinfo{volume}{70}},
  \bibinfo{pages}{062304} (\bibinfo{year}{2004}).

\bibitem[{\citenamefont{Vidal et~al.}(2001)\citenamefont{Vidal, Masanes, and
  Cirac}}]{quantph0102037}
\bibinfo{author}{\bibfnamefont{G.}~\bibnamefont{Vidal}},
  \bibinfo{author}{\bibfnamefont{L.}~\bibnamefont{Masanes}}, \bibnamefont{and}
  \bibinfo{author}{\bibfnamefont{J.}~\bibnamefont{Cirac}},
  \bibinfo{note}{e-print quant-ph/0102037} (\bibinfo{year}{2001}).

\bibitem[{\citenamefont{Nielsen and Chuang}(1997)}]{prl79_321}
\bibinfo{author}{\bibfnamefont{M.~A.} \bibnamefont{Nielsen}} \bibnamefont{and}
  \bibinfo{author}{\bibfnamefont{I.~L.} \bibnamefont{Chuang}},
  \bibinfo{journal}{Phys. Rev. Lett.} \textbf{\bibinfo{volume}{79}},
  \bibinfo{pages}{321} (\bibinfo{year}{1997}).

\bibitem[{\citenamefont{Hillery
  et~al.}(2002{\natexlab{a}})\citenamefont{Hillery, Bu{\v z}ek, and
  Ziman}}]{pra65_022301}
\bibinfo{author}{\bibfnamefont{M.}~\bibnamefont{Hillery}},
  \bibinfo{author}{\bibfnamefont{V.}~\bibnamefont{Bu{\v z}ek}},
  \bibnamefont{and} \bibinfo{author}{\bibfnamefont{M.}~\bibnamefont{Ziman}},
  \bibinfo{journal}{Phys. Rev. A} \textbf{\bibinfo{volume}{65}},
  \bibinfo{pages}{022301} (\bibinfo{year}{2002}{\natexlab{a}}).

\bibitem[{\citenamefont{Hillery
  et~al.}(2002{\natexlab{b}})\citenamefont{Hillery, Ziman, and Bu{\v
  z}ek}}]{pra66_042302}
\bibinfo{author}{\bibfnamefont{M.}~\bibnamefont{Hillery}},
  \bibinfo{author}{\bibfnamefont{M.}~\bibnamefont{Ziman}}, \bibnamefont{and}
  \bibinfo{author}{\bibfnamefont{V.}~\bibnamefont{Bu{\v z}ek}},
  \bibinfo{journal}{Phys. Rev. A} \textbf{\bibinfo{volume}{66}},
  \bibinfo{pages}{042302} (\bibinfo{year}{2002}{\natexlab{b}}).

\bibitem[{\citenamefont{Borhani and Loss}(2004)}]{quantph0410145}
\bibinfo{author}{\bibfnamefont{M.}~\bibnamefont{Borhani}} \bibnamefont{and}
  \bibinfo{author}{\bibfnamefont{D.}~\bibnamefont{Loss}},
  \bibinfo{note}{e-print quant-ph/0410145}  (\bibinfo{year}{2004}).

\bibitem[{\citenamefont{Hutton and Bose}(2004)}]{HuttonB04}
\bibinfo{author}{\bibfnamefont{A.}~\bibnamefont{Hutton}} \bibnamefont{and}
  \bibinfo{author}{\bibfnamefont{S.}~\bibnamefont{Bose}},
  \bibinfo{journal}{Phys. Rev. A} \textbf{\bibinfo{volume}{69}},
  \bibinfo{pages}{042312} (\bibinfo{year}{2004}).

\bibitem[{\citenamefont{Breuer et~al.}(2004)\citenamefont{Breuer, Burgarth, and
  Petruccione}}]{BreuerBP04}
\bibinfo{author}{\bibfnamefont{H.-P.} \bibnamefont{Breuer}},
  \bibinfo{author}{\bibfnamefont{D.}~\bibnamefont{Burgarth}}, \bibnamefont{and}
  \bibinfo{author}{\bibfnamefont{F.}~\bibnamefont{Petruccione}},
  \bibinfo{journal}{Phys. Rev. A} \textbf{\bibinfo{volume}{70}},
  \bibinfo{pages}{045323} (\bibinfo{year}{2004}).

\bibitem[{\citenamefont{Hill and Wootters}(1997)}]{HillW97}
\bibinfo{author}{\bibfnamefont{S.}~\bibnamefont{Hill}} \bibnamefont{and}
  \bibinfo{author}{\bibfnamefont{W.~K.} \bibnamefont{Wootters}},
  \bibinfo{journal}{Phys. Rev. Lett.} \textbf{\bibinfo{volume}{78}},
  \bibinfo{pages}{5022} (\bibinfo{year}{1997}).

\bibitem[{\citenamefont{Wootters}(1998)}]{Wootters98}
\bibinfo{author}{\bibfnamefont{W.~K.} \bibnamefont{Wootters}},
  \bibinfo{journal}{Phys. Rev. Lett.} \textbf{\bibinfo{volume}{80}},
  \bibinfo{pages}{2245} (\bibinfo{year}{1998}).

\bibitem[{\citenamefont{Osborne}(2005)}]{quantph0502176}
\bibinfo{author}{\bibfnamefont{T.~J.} \bibnamefont{Osborne}},
  \bibinfo{note}{e-print quant-ph/0502176} (\bibinfo{year}{2005}).

\bibitem[{\citenamefont{Bocchieri and Loinger}(1957)}]{BocchieriL57}
\bibinfo{author}{\bibfnamefont{P.}~\bibnamefont{Bocchieri}} \bibnamefont{and}
  \bibinfo{author}{\bibfnamefont{A.}~\bibnamefont{Loinger}},
  \bibinfo{journal}{Phys. Rev.} \textbf{\bibinfo{volume}{107}},
  \bibinfo{pages}{337} (\bibinfo{year}{1957}).

\end{thebibliography}

\end{document}